\documentclass[prx,aps,superscriptaddress,twocolumn,floatfix]{revtex4-2}

\usepackage{dsfont}
\usepackage{mathtext}

\usepackage{mathtools}
\usepackage[usenames,dvipsnames]{xcolor}
\usepackage{amsmath}
\usepackage{amssymb,mathrsfs}
\usepackage{graphicx}
\usepackage{epstopdf}
\usepackage{mathtext}
\usepackage{yfonts}
\usepackage{bm}

\usepackage{color}
\usepackage[colorlinks=true,linkcolor=blue]{hyperref}

\newcommand{\br}{{\bm r}}

\newcommand{\im}{{\rm im}}

\begin{document}

\title{Macroscopic Zeno effect in Su–Schrieffer–Heeger photonic topological insulator}

\author{S.~K.~Ivanov}
\email[Correspondence email address: ]{sergey.ivanov@icfo.eu}%
\affiliation{Institute of Spectroscopy, Russian Academy of Sciences, 108840, Troitsk, Moscow, Russia}

\author{S.~A.~Zhuravitskii}
\affiliation{Institute of Spectroscopy, Russian Academy of Sciences, 108840, Troitsk, Moscow, Russia}
\affiliation{Quantum Technology Centre, Faculty of Physics, M. V. Lomonosov Moscow State University, 119991, Moscow, Russia}

\author{N.~N.~Skryabin}
\affiliation{Institute of Spectroscopy, Russian Academy of Sciences, 108840, Troitsk, Moscow, Russia}
\affiliation{Quantum Technology Centre, Faculty of Physics, M. V. Lomonosov Moscow State University, 119991, Moscow, Russia}

\author{I.~V.~Dyakonov}
\affiliation{Quantum Technology Centre, Faculty of Physics, M. V. Lomonosov Moscow State University, 119991, Moscow, Russia}

\author{A.~A.~Kalinkin}
\affiliation{Institute of Spectroscopy, Russian Academy of Sciences, 108840, Troitsk, Moscow, Russia}
\affiliation{Quantum Technology Centre, Faculty of Physics, M. V. Lomonosov Moscow State University, 119991, Moscow, Russia}

\author{S.~P.~Kulik}
\affiliation{Quantum Technology Centre, Faculty of Physics, M. V. Lomonosov Moscow State University, 119991, Moscow, Russia}

\author{Y.~V.~Kartashov}
\affiliation{Institute of Spectroscopy, Russian Academy of Sciences, 108840, Troitsk, Moscow, Russia}

\author{V.~V.~Konotop}
\affiliation{Departamento de F\'isica and Centro de F\'isica Te\'orica e Computacional, Faculdade de Ci\^encias, Universidade de Lisboa, Campo Grande, Ed. C8, Lisboa 1749-016, Portugal}


\author{V.~N.~Zadkov}
\affiliation{Institute of Spectroscopy, Russian Academy of Sciences, 108840, Troitsk, Moscow, Russia}
\affiliation{Faculty of Physics, Higher School of Economics, 105066 Moscow, Russia}

\begin{abstract}

The quantum Zeno effect refers to slowing down of the decay of a quantum system that is affected by frequent measurements. Nowadays, the significance of this paradigm is extended far beyond quantum systems, where it was introduced, finding physical and mathematical analogies in such phenomena as the suppression of output beam decay by sufficiently strong absorption introduced in guiding optical systems. In the latter case, the effect is often termed as macroscopic Zeno effect. Recent studies in optics, where enhanced transparency of the entire system was observed upon the increase of the absorption, were largely focused on the systems obeying parity-time symmetry, hence, the observed effect was attributed to the symmetry breaking. While manifesting certain similarities in the behavior of the transparency of the system with the mentioned studies, the macroscopic Zeno phenomenon reported here in topological photonic system is far more general in nature. In particular, we show that it does not require the existence of exceptional points, and that it is based on the suppression of decay for only a subspace of modes that can propagate in the system, alike the quantum Zeno dynamics. By introducing controlled losses in one of the arms of a topological insulator comprising two closely positioned Su-Schrieffer-Heeger arrays, we demonstrate the macroscopic Zeno effect, which manifests itself in an increase of the transparency of the system with respect to the topological modes created at the interface between two arrays. The phenomenon remains robust against disorder in the non-Hermitian topological regime. In contrast, coupling a topological array with a non-topological one results in a monotonic decrease in output power with increasing absorption. This research broadens the understanding of the Zeno effect and illustrates its potential for control of light propagation, switching and steering in non-Hermitian topological systems.


\end{abstract}

\maketitle

\section{Introduction}\label{sec1}
The phenomenon of topological insulation is a rich source of new fundamental effects and potential applications utilizing the robustness of topological edge states~\cite{Hafezi-11,Harari-18,Bandres-18,Guglielmon-19} resulting in a broad interdisciplinary concept covering diverse areas of science~\cite{HasanKane-10,QiZhang-11,CooperDalibardSpielman-19}.
Topologically nontrivial systems are renowned for their distinctive and exceptional robustness against local deformations, which makes them  conceptually perfect settings for the implementation of various routing and switching schemes. In optics, various switching and coupling mechanisms have been explored, such as Floquet insulators realized with helical waveguide arrays~\cite{BazhanMalomed-21} and Su‐Schrieffer‐Heeger chains~\cite{SongSun-20,Efremidis-21,ArkhipovaIvanov-22}. The resonant switching of topologically protected edge states and valley Hall edge states have been theoretically studied in~\cite{ZhangKartashov-18,ZhongKartashov-19}.
{{The studies conducted in~\cite{ZhangChen-19,IvanovKartashov-20} investigate the resonant coupling between two topological edge states occurring at the interface between two arrays.}}
The switching of excitations between the opposite edges of a topologically nontrivial sample under pumping was experimentally observed in~\cite{KrausLahini-12,ZilberbergHuang-18}. The discovery of many intriguing phenomena in topological physics was made with the use of dissipative non-Hermitian systems~\cite{RudnerLevitov-09,DiehlRico-11,EsakiSato-11,BardynBaranov-13,Schomerus-13,Yuce-15}, which  provide a plethora of alternative designs for controlling light~\cite{GuoSalamo-09,RuterMakris-10,RegensburgerBersch-12,GanainyMakris-18}. Progress here has led to the experimental observation of topological non-Hermitian edge~\cite{PoliBellec-15,WeimannKremer-17,ZeunerRechtsman-15} and bulk~\cite{ZeunerRechtsman-15} states, topological light steering~\cite{ZhaoQiao-19}, funneling~\cite{WeidemannKremer-19} and topological insulator lasing~\cite{Harari-18,Bandres-18,BahariNdao-17,KartashovSkryabin-19,IvanovZhang-19}.

By and large, the exploding development of topological optics can be explained by the mathematical similarities of the optical models and the Schr\"odinger equation governing   electrons in solids. Such analogy extends to both Hermitian and non-Hermitian settings. An interesting fundamental phenomenon that can be analyzed from this perspective is the quantum Zeno effect. This phenomenon, introduced first in quantum physics, reveals in the strong suppression of the decay of a quantum system subjected to frequent measurements. The quantum Zeno effect was predicted theoretically as a counter-intuitive phenomenon~\cite{MisraSudarshan-77,Peres-80} (see also \cite{FacchiPascazio-08}) and its subsequent experimental demonstration has triggered extensive  fundamental~\cite{ItanoHeinzen-90,FischerGutierrez-01,BaroneKurizki-04,SyassenBauer-08,KalbCramer-16}
and applied \cite{BeigeBraun-00,FacchiHradil-02,Paz-SilvaRezakhani-12,Paz-SilvaRezakhani-12,BarontiniHohmann-15} studies.
Nowadays, it is widely understood that the Zeno effect is a ubiquitous phenomenon, extending far beyond quantum systems and that it can be viewed as a macroscopic dissipative phenomenon, hence the term the {\em macroscopic Zeno effect}. An analogy between the quantum Zeno effect and the decay of light in an array of optical waveguides was shown in~\cite{Longhi-06}. A general analogy between the meanfield description of a frequently measured quantum system and a dissipative macroscopic system emulated by a dissipative dimer was shown in~\cite{ShchesnovichKonotop-10}. Optical realizations of the macroscopic Zeno effect using two coupled waveguides, one of which is lossy, were proposed in~\cite{AbdullaevKonotopShchesnovich-11,AbdullaevKonotopOgren-11,ThapliyalPathak-16}. The macroscopic Zeno effect has been observed also with the light propagating in a waveguide tunnel-coupled to the semi-infinite array~\cite{BiagioniDella-08}.
 

{{In this study, we experimentally demonstrate this effect in a topological non-Hermitian system consisting of two finite Su-Schrieffer-Heeger (SSH)~\cite{SuSchrieffer-79} waveguide arrays, where controllable dissipation is introduced in one of the arrays. The SSH arrays comprise evanescently coupled waveguides placed in close proximity along a line. This work has three primary objectives. First, we experimentally demonstrate the macroscopic Zeno effect that is characterized by a counter-intuitive enhancement of the transparency of the whole sample with the increase of losses in a part of it. Second, we show that the occurrence of the macroscopic Zeno effect crucially depends on whether the system is in topological or in trivial phase. When both SSH arrays are in topological phases, the presence of {\em two coupled topological edge states} enables light switching between them. Introducing controllable linear losses in one of the arrays we observe the optical Zeno effect for such topological edge states, that coexist with delocalized bulk modes in this complex system. On the other hand, the Zeno effect is not observed when the array in the topological phase is placed in contact with the array in a trivial phase, since in this case switching between arrays is inhibited and introduced losses lead only to monotonic decrease of the transparency of the system. Third, we demonstrate that the control and even complete arrest of switching between topological arrays can be achieved by increasing losses in one of the arrays. This indicates that non-Hermitian topological systems can support controllable robust transport processes even under the extreme complexity introduced by dissipation, an important feature for contemporary topological applications and for design of new compact switching devices.}}

{{It must be stressed that even though some features of switching dynamics in our setting may look similar to those in a passive parity-time ($\mathcal{PT}$) dimer explored in previous experiments, where the $\mathcal{PT}-$symmetry breaking was observed~\cite{GuoSalamo-09} and where reduction of losses for the output beam was related to the existence of an exceptional point (EP) separating different phases of the system, the topological system studied here and the macroscopic Zeno effect itself are conceptually different from the above mentioned effects based on the $\mathcal{PT}$ symmetry. Indeed, as will be shown below, our system does not obey the $\mathcal{PT}$ symmetry (even passive one) and its spectrum does not have the EPs. Unlike in Ref.~\cite{GuoSalamo-09}, the Zeno effect reported here is observed only in a topologically nontrivial phase for a subspace of the edge states, rather than for the whole space of the modes guided by the system. Respectively, the mathematical structure and physical background of a minimal two-mode model that can approximately describe dynamics of our system is completely different from the model for two coupled waveguides used in Ref.~\cite{GuoSalamo-09}.}}


\section{Results}
\subsection{Theory}

\begin{figure*}[t]
\centering
\includegraphics[width=1\textwidth]{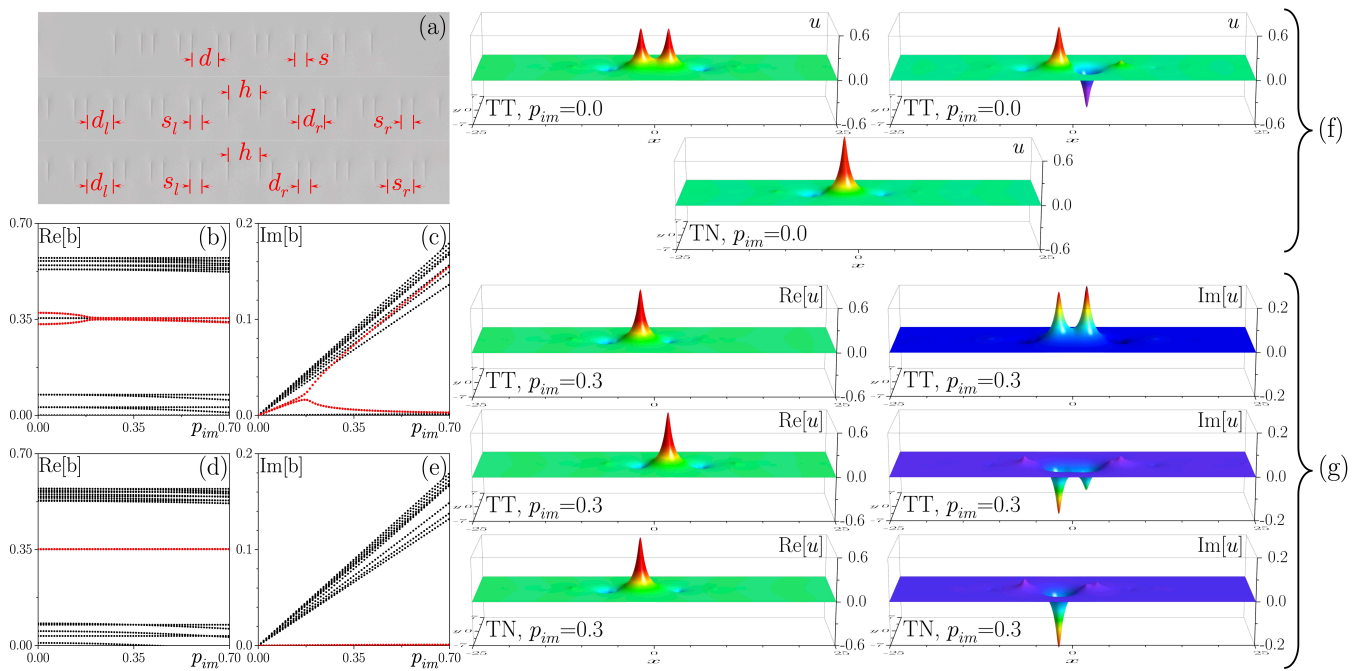}
\caption{
(a) Microphotographs of the SSH array in the topological phase with $d=33$~$\mu m$ and $s=15$~$\mu m$ (top row), two SSH arrays in the TT configuration with $d_l=d_r= 33$~$\mu m$, $s_l=s_r = 15$~$\mu m$ (middle row), two SSH arrays in the TN configuration with $d_l=s_r= 33~\mu m$, $s_l=d_r = 15~\mu m$ (bottom row). (b), (c) Real and imaginary parts of the propagation constant $b$ for the TT configuration as a function of the imaginary depth $p_{\rm im}$ of the right waveguides array. (d), (e) Real and imaginary parts of the propagation constant for the TN configuration as a function of $p_{\rm im}$. Red dots correspond to the topological states localized at the interface between the arrays. (f) Profiles $u(\br)$ of the modes forming at the interface between two arrays for the TT configuration (top row) and TN configuration (bottom row) for $p_{\rm im}=0$. (g) Real (left column) and imaginary (right column) parts of the mode profiles for the TT configuration (top and middle rows show two existing modes) and the TN configuration (bottom row) for $p_{\rm im}=0.3$. Different color scales are used for the real and imaginary parts of $u$. In all cases, $h=39$~$\mu m$. Here and below $p_{\rm re}=4.7$ ($\delta n \approx 5.3\times10^{-4}$), $\sigma_x=2.5$~$\mu m$, and $\sigma_y=7.5$~$\mu m$.}
\label{fig1}
\end{figure*}

We start with the theoretical description of the topological states in the dissipative SSH lattices shown in~Fig.~\ref{fig1}(a). Under the paraxial approximation, the dimensionless light field amplitude $\psi$ is governed by the Schr\"{o}dinger equation
\begin{equation} 
\label{eqNLS}
    i\frac{\partial\psi}{\partial z}= H (\br)\psi, 
\end{equation}

\noindent where ${\bm r}=(x,y)$ are the transverse coordinates normalized to the characteristic transverse scale $w_0=10$~$\mu \textrm{m}$, $z$ is the propagation distance scaled to the diffraction length $\kappa w_0^2$, $\kappa=2\pi n_r/\lambda$ is the wavenumber, $\lambda=800$~$\textrm{nm}$ is the wavelength, $n_r\approx 1.45$ is the real part of the unperturbed refractive index of the fused silica glass, where our structures are inscribed, $\nabla=(\partial_x,\partial_y)$, and the Hamiltonian is given by 
\begin{align}
    \label{H0}
     H(\br)\equiv -\frac{1}{2}\nabla^2 
     -V(\br)-iW(\br).
\end{align}

\noindent The real part of the optical potential, $V(\br)$, describes two SSH arrays of the waveguides placed along the $x-$axis:
\begin{align}
\label{potential}
 V(\br) = \!\! \sum_{n=-\infty}^{-1} V_0(x-x_n^{\rm l},y-y_n^{\rm l})+\sum_{n=0}^{\infty} V_0(x-x_n^{\rm r},y-y_n^{\rm r}),
\end{align}

\noindent where $ V_0(x,y)=   p_{\rm re}\exp\left(-{x^2}/{\sigma_x^2}-{y^2}/{\sigma_y^2} \right)$, $x_n^{\rm l,r}$ and $y_n^{\rm l,r}$ are the nodes of the left (l) and right (r) SSH grids, and $\sigma_{x,y}$ are the widths of the individual waveguides (they are elliptical due to the fs-writing procedure, i.e., $\sigma_x\neq\sigma_y$) along the $x$ and $y$ axes. The index $n$ in (\ref{potential}) enumerates waveguides in the left ($n<0$) and  right  ($n\geq 0$) arrays. In the dimensionless units the modulation depth is $p_{\rm re}=\kappa^2 w_0^2 \delta n_r/n_r$ with $\delta n_r$ being the refractive index contrast. Spatially inhomogeneous losses are present only in the right array, i.e., 
\begin{align}
    W(\br) = \Gamma \sum_{n=0}^{\infty} V_0(x-x_n^{\rm r},y-y_n^{\rm r}), \qquad \Gamma=\frac{p_{\im}}{p_{\rm re}}.
\end{align} 

\noindent The depth of the imaginary part of the potential $p_{\rm im}$, i.e., $\Gamma$ is considered as a control parameter.

In order to introduce a topological phase into this structure, it is necessary to tune intra-cell and inter-cell couplings in each unit cell (containing two waveguides)~\cite{SuSchrieffer-79}. This can be done by introducing a shift of the waveguides (in the opposite directions for two waveguides in the unit cell), so that the inter-cell waveguide spacing $s$ becomes smaller than the intra-cell spacing $d$  [see upper microscopic image of the array with 7 cells in Fig.~\ref{fig1}(a), where $d$ and $s$ are indicated]. At $d=s$ the array represents an ordinary periodic structure with equal distance between all waveguides. The truncated array shown in the upper line in Fig~\ref{fig1}(a) is topological if $s < d$. Then one observes the appearance of the topological states localized at the array edges with eigenvalues within the topological gap. Otherwise, at $s > d$, the array is in the trivial phase and no edge states can be formed in the trivial gap in the spectrum. 

We further place two such arrays into close proximity and denote the distance between them as $h$. This distance is varied in the experiments. Such a setup is illustrated in the middle and bottom rows of Fig.~\ref{fig1}(a). The first case (denoted below as TT) shows the  configuration where $s_{l,r} < d_{l,r}$ and both arrays are topologically nontrivial. In the second case shown in the bottom line of Fig.~\ref{fig1}(a) (denoted below by TN) the left array has the topological  ($s_{l} < d_{l}$) configuration while the right array is in the topologically trivial phase ($s_r > d_r$). Below we will show that only the TT configuration supports light switching between two arrays, provided that the right array is lossy. 

Based on this concept, we put our focus on the analysis of the dissipative system. The imaginary part of the potential, $W(\br)$, describes losses introduced only into the right array in our structure, i.e. we assume that the left array is transparent. Parameter $p_{\rm im}$ controls the amount of losses. We numerically calculated the transformation of the spectrum for the TT and TN configurations with increase of $p_{\rm im}$. The eigenmodes for Schr\"{o}dinger equation can be written in the form $\psi(\br,z)=u(\br) e^{ibz}$, where $b$ is the eigenvalue (propagation constant) of the mode and function $u$ describes the mode profile. For a conservative system, the propagation constant $b$ is real, while for a dissipative one $b$ may be complex. The dependence of the real and imaginary parts of $b$ on $p_{\rm im}$ for the TT configuration is shown in Figs.~\ref{fig1}(b) and~(c) for the array parameters close to the experimental ones, i.e. for $p_{\rm re}=4.7$ ($\delta n \approx 5.3\times10^{-4}$), $d_l=d_r= 3.3$ ($33$~$\mu m$), $s_l=s_r = 1.5$ ($15$~$\mu m$), $\sigma_x = 0.25$ ($2.5$~$\mu m$), and $\sigma_y = 0.75$ ($7.5$~$\mu m$). Red dots correspond to the modes localized at the interface between two topological arrays. As can be seen, these branches belong to the topological gap, which indicates that the structure is in the topological phase for a sufficiently small $p_{\rm im}$ (SSH topological insulator with non-Hermitian domain wall was also discussed theoretically in~\cite{LiFan-22}). Modes of the other two branches within the gap [black dots close to the red ones in Fig.~\ref{fig1}(b)] are localized at the outer edges of the entire structure and are not considered here. In what follows, we will consider only modes localized at the interface. The spectrum of the TN configuration is illustrated in Figs.~\ref{fig1}(d) and~(e) for $d_l=s_r= 3.3$ ($33~\mu m$), $s_l=d_r = 1.5$ ($15~\mu m$). The topological gap has only one edge level (red dots), corresponding to the mode localized on the right edge of the left array. {{When $s_{l,r} > d_{l,r}$, both arrays are in trivial phase and no edge states can form within the trivial gap in the spectrum (see Supporting Information).}}

The topological eigenstates for different configurations and magnitude of losses are shown on the right side of Fig.~\ref{fig1}, where Fig.~\ref{fig1}(f) corresponds to the conservative TT (top row) and the TN (bottom row) configurations and $u$ is the real function. One can see that in the case of TT configuration there exist two modes (in-phase and out-of-phase)  residing at the interface between the arrays. Excitation of the edge channel only on the left or right side of the interface is equivalent to the simultaneous excitation of the in-phase and out-of-phase modes with nearly equal weights. This type of excitation initiates switching between two subsystems, where the beating length is inversely proportional to the difference in propagation constants $Z_s=\pi/(b_1-b_2)$. $Z_s$ increases with the increase of the distance between the arrays $h$ due to diminishing of the propagation constant difference $b_1-b_2$ (see, e.g.,~\cite{ArkhipovaIvanov-22}). Conservative TN configuration, in turn, supports only a single topological mode on the rightmost edge of the left array. Accordingly, the single channel excitation does not cause switching in such a structure. In the case of a dissipative system, the profiles of eigenmodes are no longer remain purely real functions. Fig.~\ref{fig1}(g) shows the distribution of the real (left column) and imaginary (right column) parts of the function $u$ at $p_{\rm im}=0.3$ for both TT and TN configurations. In TT configuration, two topological modes at the interface between two arrays become notably asymmetric and gradually concentrate at the different sides of the interface with the increase of $p_{\rm im}$. It should be noted that although $u$ does not depend on $z$, nevertheless the excitation of such modes is accompanied by the energy losses due to the dissipation present in the right array. Meantime, as we will show,  the effective losses of the entire system can be achieved by increasing the losses in the right array. This is the manifestation of the macroscopic Zeno phenomenon.

\subsection{Two-mode model}

A direct insight on the physics of the macroscopic Zeno effect can be gained using the two-mode model. Here we describe it for the TT configuration using the advantage that this phenomenon is based on the interconnected evolution of two interface states (topological modes), which are {strongly concentrated on the right outermost waveguide of the conservative array and left outermost waveguide of the dissipative array}. Denoting the respective complex field amplitudes by $A_0$ and $B_0$ we obtain the two-mode model (see Supporting Information for the derivation):
\begin{align}
\label{two-modes}
	\begin{array}{l}
		\displaystyle -i\frac{dA_0}{dz} =iw_a A_0 +v_h(1+i\Gamma )B_0,
		\\[2mm]
		\displaystyle -i\frac{dB_0}{dz} =iw_b B_0 +v_h(1+i\Gamma) A_0, 
	\end{array}
\end{align}

\noindent where
\begin{align*}
	w_a=& \Gamma\int_{\mathbb{R}^2}|\phi(\br)|^2V_0(x-h,y)dxdy, 
 	\\
 	w_b=& \Gamma\int_{\mathbb{R}^2}|\phi(\br)|^2V_0(x,y)dxdy,   
 	 \\
 	v_h =&\int_{\mathbb{R}^2}\phi^*(\br) V_0(x - h,y)\phi(\br-h{\bf i})dxdy.
\end{align*}

\noindent Here $\textbf{i}$ is the unit vector along $x-$axis, and $\phi(\br)$ is the fundamental mode of the potential $V_0(x,y)$ with the largest propagation constant $\beta$:
 \begin{align}
 	\label{H00}
 	 H_0(\br)\phi(\br)=\beta \phi(\br),
 	  \quad H(\br) \equiv -\frac{1}{2}\nabla^2 - V_0(x,y). 
 \end{align}

\begin{figure}[t]
\centering
\includegraphics[width=1.00\linewidth]{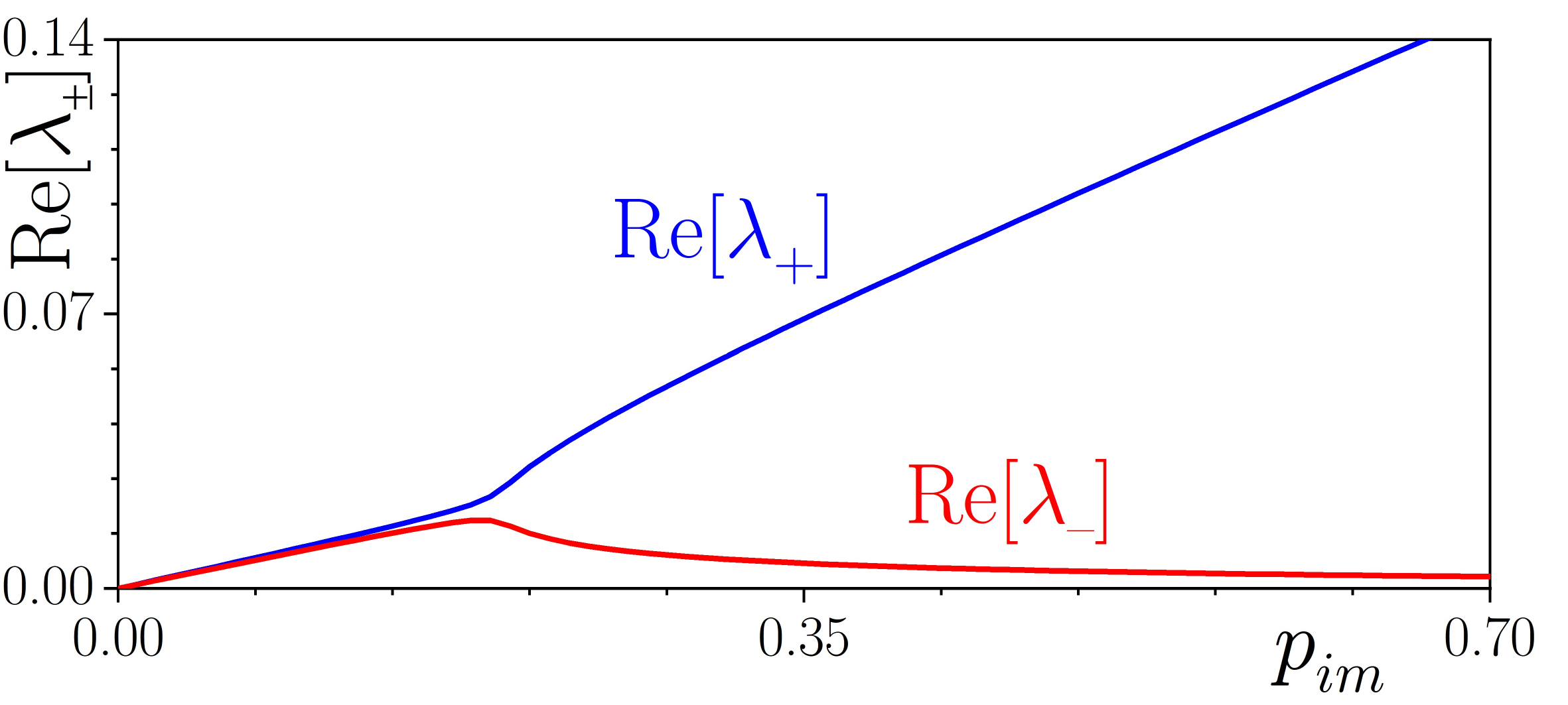}
\caption{
Real part of $\lambda_+$ (blue) and $\lambda_-$ (red) versus $p_{\rm im}$ are shown. Here $h=39$~$\mu m$, $w_a=0.0001p_{\rm im}$, $w_b=0.2146p_{\rm im}$, and $v_h=0.021$.}
\label{fig2}
\end{figure}

The system (\ref{two-modes}) is readily solved by the ansatz $A_0,B_0\propto  e^{-\lambda z}$, yielding two complex decrements
\begin{align}
\label{lambda}
\lambda_\pm=\frac{1}{2}(w_a+w_b\pm \Lambda^{1/2}), \qquad \Lambda=(w_a-w_b)^2+4v_h^2(\Gamma-i)^2
\end{align}
{{with the corresponding (non-normalized) eigenvectors
\begin{align}
    C_\pm=\left(
     \begin{matrix}
    2v_h(\Gamma-i)
        \\
       w_b-w_a\pm\sqrt{\Lambda}
    \end{matrix}\right).
\end{align}
}}

\noindent In Fig.~\ref{fig2} we plot Re$(\lambda_\pm)$ for the parameters used in the experiment with $h=39$~$\mu \rm m$. It follows from this dependence that while the decay rate for one of the modes monotonically increases with the increase of the parameter $p_{\rm{im}}$ quantifying losses in the right array, for another mode it instead may decrease with $p_{\rm{im}}$ representing the manifestation of the Zeno effect. This dependence closely resembles the evolution of the imaginary part $\textrm{Im}(b)$ of the propagation constant of two topological edge modes with $p_{\rm{im}}$ presented with the red dots in Fig.~\ref{fig1}(c), demonstrating the fact that the effective absorption for one of these modes may actually decrease with the increase of the losses in the right array. This result illustrates the absence of an EP in the system: the blue and red lines in Fig.~\ref{fig2} do not cross for all $p_{\rm{im}}>0$, even near the maximum of the red line {{where the behavior of the curves resembles behavior in a vicinity of an EP. Mathematically, however, EP can exist only if $\Lambda=0$ in (\ref{lambda}): then both eigenvalues $\lambda_\pm$ and eigenvectors $C_\pm$ coalesce. However,  $\Lambda\neq 0$  for all $v_h\neq 0$  (recall that all of the parameters in (\ref{lambda}) are real). On the other hand, if $v_h=0$ , the system (\ref{two-modes}) becomes decoupled. Thus, there are no EPs in this system.}}


\begin{figure}[t]
\centering
\includegraphics[width=1.00\linewidth]{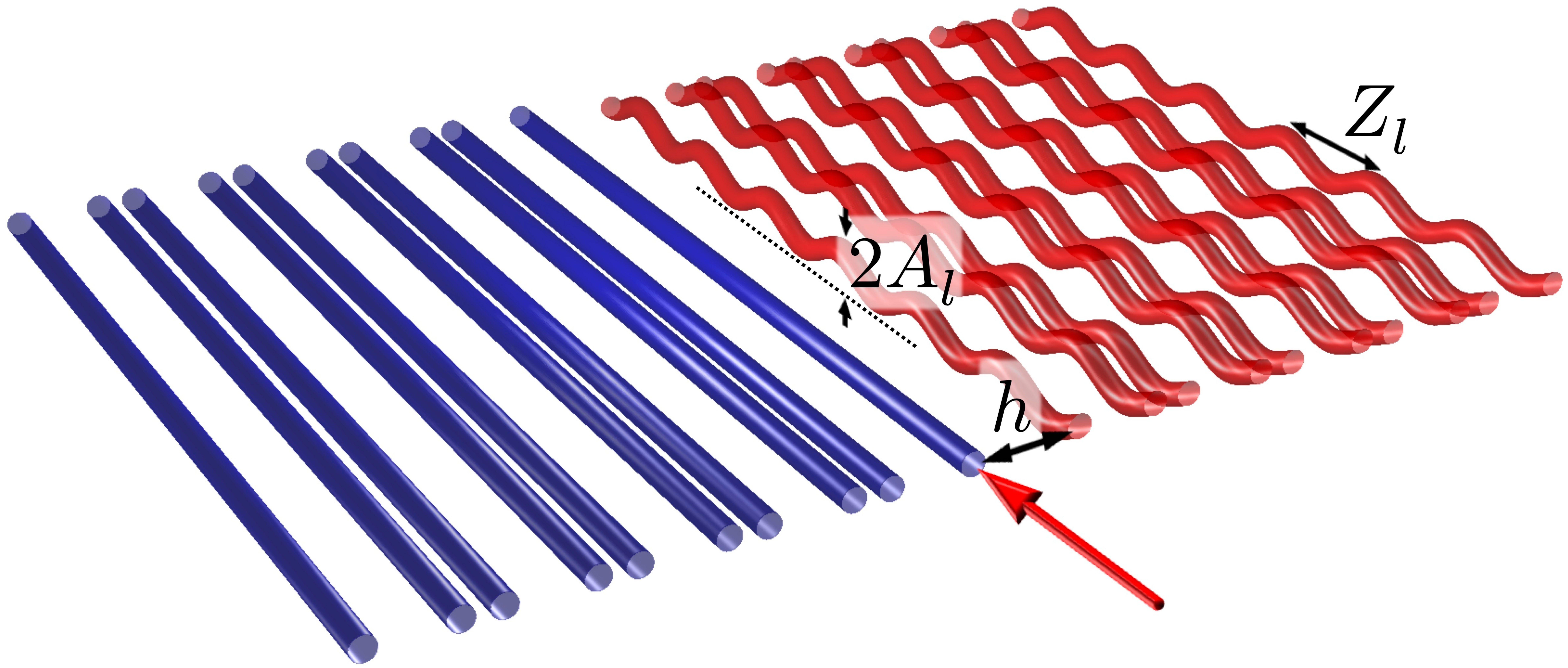}
\caption{
Schematic representation of the interface between two topological arrays where the red arrow indicates the excited waveguide in the left array (marked in blue) with straight channels. The sites in the right array (marked in red) oscillate in the vertical direction with the period $Z_l$ and the amplitude $A_l$, causing bending or radiation losses to the continuum modes.}
\label{fig3}
\end{figure}

\begin{figure}[t]
\centering
\includegraphics[width=1.00\linewidth]{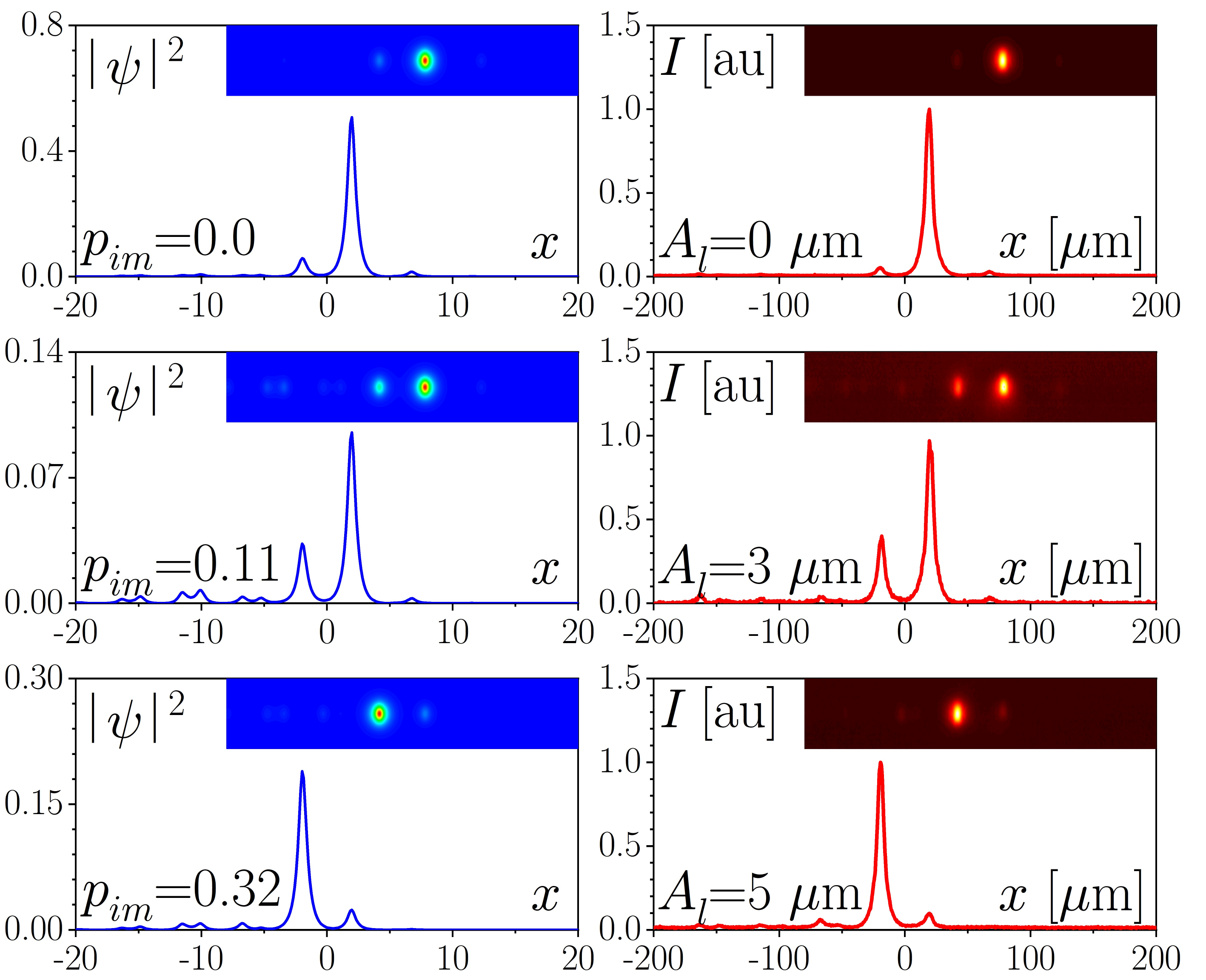}
\caption{
Cross-sections at $y=0$ of the intensity distributions after $10$~$\rm cm$ of propagation through the sample for the different losses $\gamma=0$~$\rm cm^{-1}$ (top row), $\gamma=0.43$~$\rm cm^{-1}$ (middle row), and $\gamma=1.18$~$\rm cm^{-1}$ (bottom row) introduced in the right arm. The left column shows theoretical {{results in dimensionless variables}}, while the right column shows experimental distributions. Inserts demonstrate two-dimensional output intensity distributions. Here $h=39$~$\mu{\rm m}$.}
\label{fig4}
\end{figure}

\begin{figure}[t]
\centering
\includegraphics[width=0.90\linewidth]{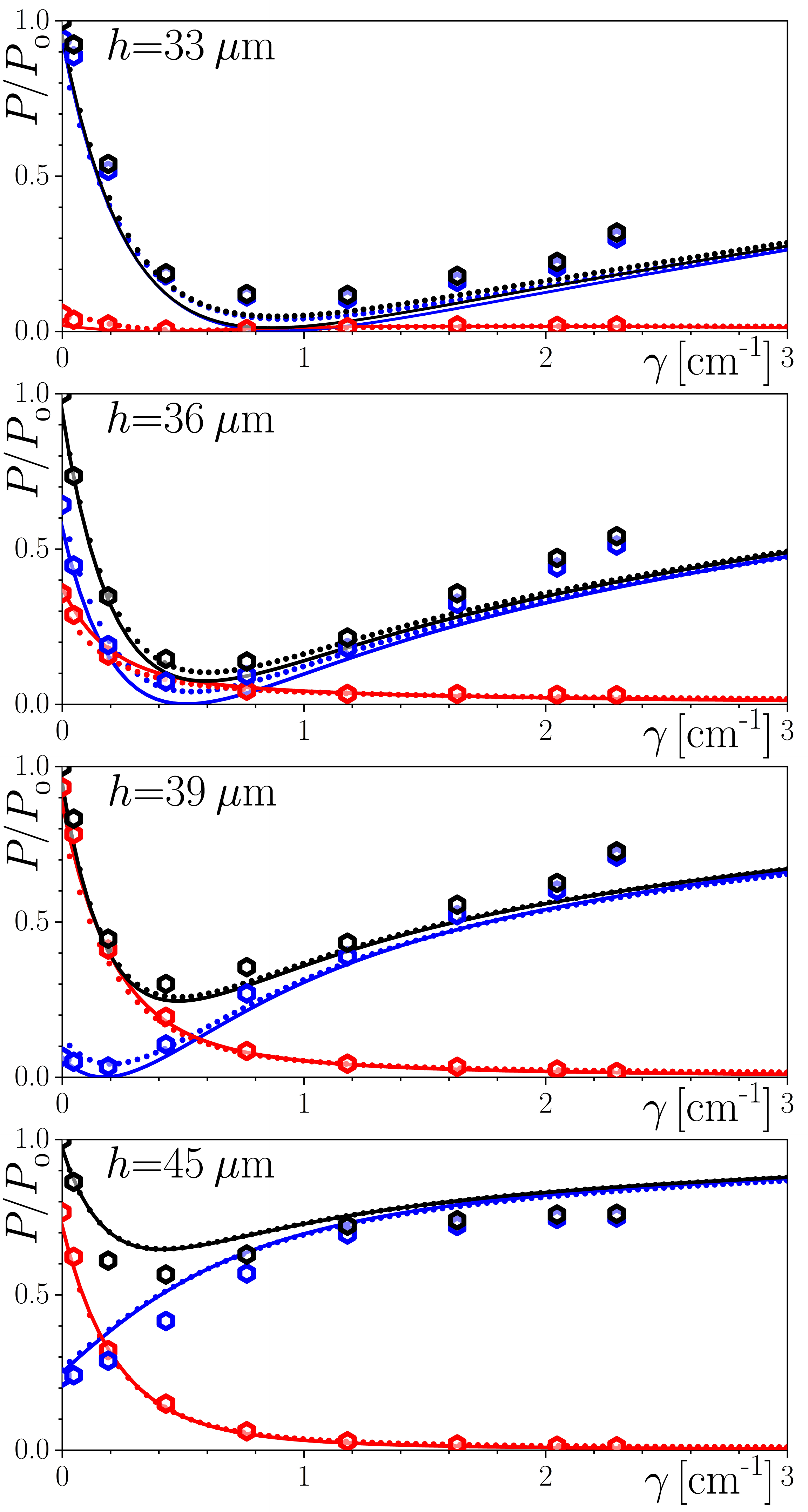}
\caption{
Dependencies of the output power $P$ on the loss factor $\gamma$ for the different distances between the arrays $h$ for the TT configuration. The power is normalized to the output power of the sample with straight waveguides $P_0$ (without additional losses in the right array). The hexagons show the experimental results, the solid curves correspond to the two-mode model, and the small circles show the results of the numerical calculations. Black, blue, and red colors correspond to the total output power, and the output power in the left (lossless) and right (lossy) arrays, respectively.}
\label{fig5}
\end{figure}

\subsection{Experimental Results}

For the experimental observation of the macroscopic Zeno effect we used the SSH arrays with $20$ waveguides ($5$ dimers in each array) inscribed in $10$~$\rm cm$-long fused silica glass samples. We determined a specific value of $p_{\rm re}$ and $\sigma_{x,y}$ using test arrays of straight waveguides. Such straight waveguides exhibit about $0.07$~$\rm cm^{-1}$ propagation losses at the wavelength of $800$~$\rm nm$ used in the experiments, which can be factored out because they are identical in every waveguide of the structure. To introduce controllable extra losses in our experimental setup, we implemented a rapid $z$-wiggling of each waveguide in the right array that causes additional radiative losses (coupling to the continuum states) due to the curvature of the guides.
This technique of creation of controllable inhomogeneous losses was introduced in ~\cite{ZeunerRechtsman-15,WeimannKremer-17,EichelkrautHeilmann-13,EichelkrautWeimann-14}.
A schematic diagram of the TT configuration with the oscillations of the waveguide in the right array is depicted in Fig.~\ref{fig3}.
The wiggling (cosine-like) occurs in the $y$-direction, transverse to the array axis, and is characterized by the oscillation period $Z_l\gg\kappa w_0^2$ and relatively small amplitude $A_l$. In our experiment, we fix the oscillation period $Z_l=0.6$ $\rm cm$ and vary the amplitude $A_l$. For the range of wiggling amplitudes used in our work (from $0$ to $8$~$\mu \rm m$), an increase in $A_l$ entails an increase in effective loss factor $\gamma$ in each single waveguide (see Methods). It is important to emphasize that although the waveguide oscillations lead to periodic change of the distance between two arrays, for wiggling amplitudes considered here this change is very small compared to $h$, and according to our numerical and experimental estimations, it can change the output power distribution in the subsystems by no more than $5\%$.


Introduction of losses into the right array has a dramatic effect on the dynamics of light switching in our structure. To illustrate the phenomenon, we use the TT configuration with a distance between the arrays of $h=39$~$\mu\rm m$. In the conservative limit (i.e., in the absence of wiggling), one observes in such a structure a complete switching of the light from one array to another one over the sample length, i.e., the length of the sample approximately corresponds to the beating length $Z_s$ for the selected $h$ value. For example, in the right column of Fig.~\ref{fig4} we show the experimental output intensity cross-sections at $y=0$ observed for the single-site excitation of the right outermost waveguide in the left array at $z=0$ for the different level of losses in the right array starting from the conservative case $\gamma=0$. One can see that while in the conservative case light fully switches into the right array, with increasing losses in the right array (i.e. with increasing the amplitude $A_l$ of the waveguide wiggling in this array, see Methods for connection between $\gamma$ and the wiggling amplitude) the energy at the output redistributes between both arrays and for sufficiently large losses $\gamma$ switching is nearly completely arrested and the light remains in the left conservative array. Further increase of $\gamma$ does not essentially change the output field distribution. This picture is reproduced in the theoretical analysis, where we simulate the light propagation in this structure using Eq. (\ref{eqNLS}) for gradually increasing values of the imaginary depth $p_{\rm im}$ defining the loss factor $\gamma$ in the right array (see Methods). Numerical results confirming the gradual arrest of switching by the losses are presented in the left column of Fig.~\ref{fig4}, {where we used input beam launched into the outermost waveguide of the left array, whose amplitude is normalized such that at $z=0$ one has $\int |\psi|^2dxdy=1$}. Insets in the experimental and theoretical panels illustrate the two-dimensional output intensity distributions. Similar behavior was observed experimentally for various distances $h$. This indicates that not only switching dynamics, but also overall losses in the system may be controlled by $\gamma$.


To characterize the manifestation of the Zeno effect in our topological system we experimentally measure the dependence of the output total power $P$
as well as the output power concentrated in the left $P_l$ 
and right $P_r$
arrays on the strength of losses in the right array for the TT configuration. We normalize the power to the output power of the system with straight waveguides $P_0$ in order to not take into account the effect of homogeneous sample losses. The dependencies of powers $P,P_l,P_r$ on $\gamma$ are shown in Fig.~\ref{fig5} for different distances $h$ between arrays, where large hexagons show the experimental results, while small dots illustrate the results of numerical simulations based on Eq. (\ref{eqNLS}). As one can see from this figure, the total power (black dots) at the output face of the sample first decreases with $\gamma$, but then starts growing when $\gamma$ exceeds the certain critical value---this is the direct manifestation of the Zeno effect. From Fig.~\ref{fig5} one can conclude that the emergence of the phenomenon is associated with the interplay between the coupling of the subsystems (arrays) and dissipation. The smaller the coupling between arrays (i.e., the larger is $h$), the smaller the amount of losses that are needed to transfer the system into the loss-induced transparency regime.

A remarkable feature observable in all panels of Fig.~\ref{fig5} is a very accurate prediction of the phenomenon by the simplified two-mode model (\ref{two-modes}), in spite of the distributed character of the system. This is another manifestation of the topological properties of the system that lead to strong localization of light at the interface between two arrays. Using the analogy with quantum physics one can identify these modes as a Hilbert space, where the Zeno phenomenon is observed. A natural consequence of this interpretation is that macroscopic Zeno phenomenon cannot be observed when one of the arrays has trivial topology [the lowest array in Fig.~\ref{fig1}(a)], since the respective Hilbert space is reduced to only one edge mode.

In terms of the spectrum, the observed effect can be understood as a transition from the increase to decrease of the imaginary part of the propagation constant for one of the red branches associated with the topological modes residing at the interface [see the spectrum in Fig.~\ref{fig1}(c)]. For instance, this transition in the modal spectrum for the distance $h=39$~$\mu\rm m$ occurs approximately at $p_{\rm im}=0.17$ (corresponding to the loss coefficient of $\gamma=0.6$~$\rm cm^{-1}$), which is in a reasonable agreement with the experimental results.



\begin{figure}[t]
\centering
\includegraphics[width=0.90\linewidth]{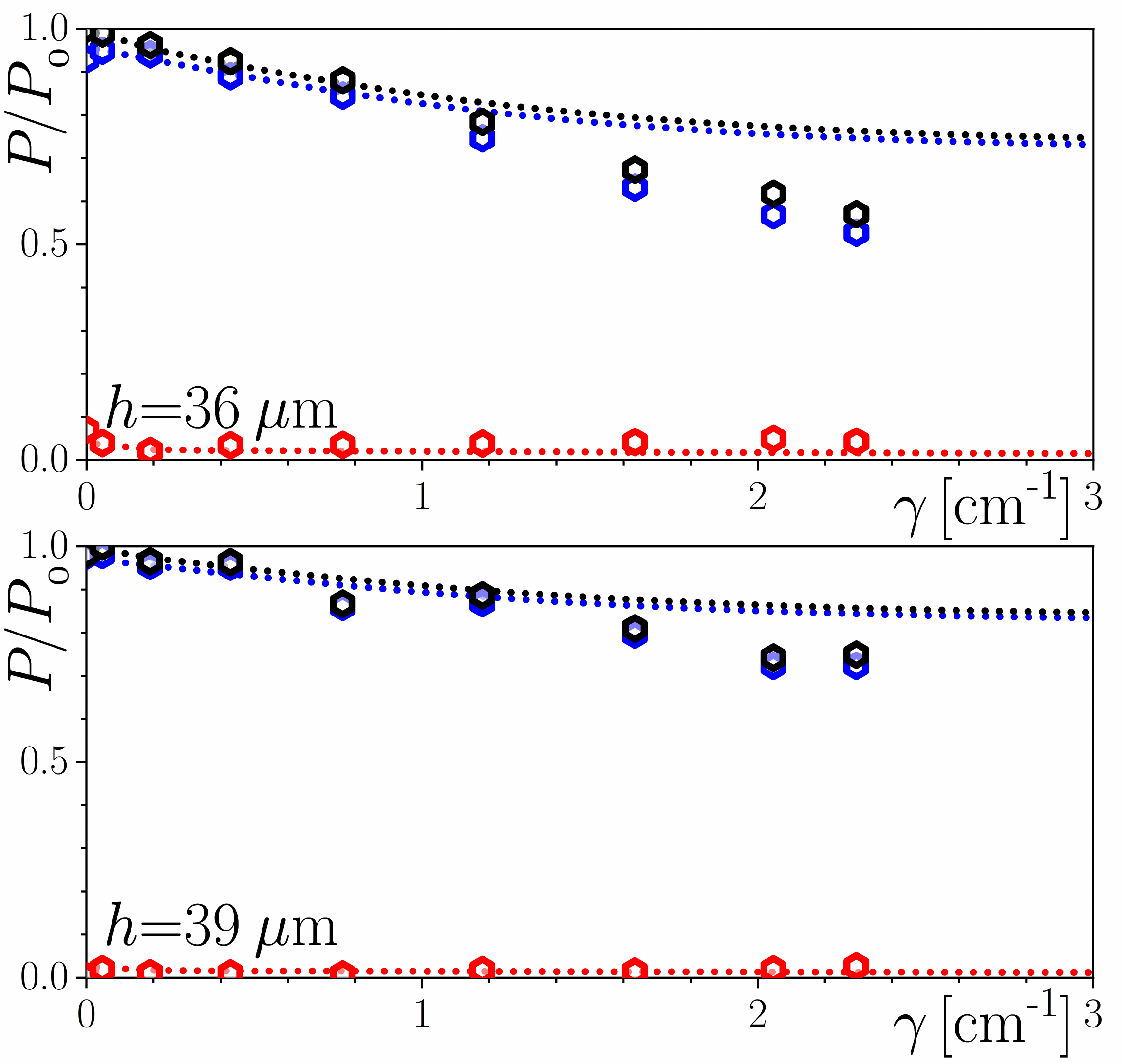}
\caption{
Dependencies of the output power $P$ and powers in the left $P_l$ and right $P_r$ arrays on the loss factor $\gamma$ for different distances between the arrays $h$ for the TN configuration. The power is normalized to the output power of the sample with straight waveguides $P_0$ (without additional losses in the right array). Hexagons show the experimental results, while small circles show the results of the numerical calculations. Black, blue, and red colors correspond to the total output power, the output power in the left (lossless) and right (lossy) arrays, respectively.}
\label{fig6}
\end{figure}

It is also essential to demonstrate the influence of the topological phase transition in one of the arrays on the observation of the Zeno effect. To this end, we now put in close proximity the topological left and nontopological right arrays, i.e., we consider the TN configuration illustrated in the bottom row of Fig.~\ref{fig1}(a). As noted above, in this case, the excitation of the outermost right waveguide of the left transparent array does not lead to the light switching.  Figure~\ref{fig6} shows the experimentally measured and numerically calculated dependencies of the normalized output power on the losses $\gamma$ for the TN case. It can be readily observed that, in contrast with the TT configuration, the Zeno effect does not occur now. {{We have also investigated the impact of losses on the dynamics of the system in a configuration where both arrays are in the trivial phase. In this NN case, we excited the outermost waveguide of the left array while varying the losses in the right array. Due to the absence of localized edge states in this case, light dramatically spreads in this regime, mainly in the left array. In this case, the power shows slightly nonmonotonic behaviour with the increase of losses, see Supporting Information. Notice however, that in this case the variation of the output power is very weak in comparison with the results of Fig. \ref{fig5} for the TT configuration because the amount of power coupling to the right array is rather small.}}

\subsection{Robustness of the effect}

\begin{figure}[t]
\centering
\includegraphics[width=1.0\linewidth]{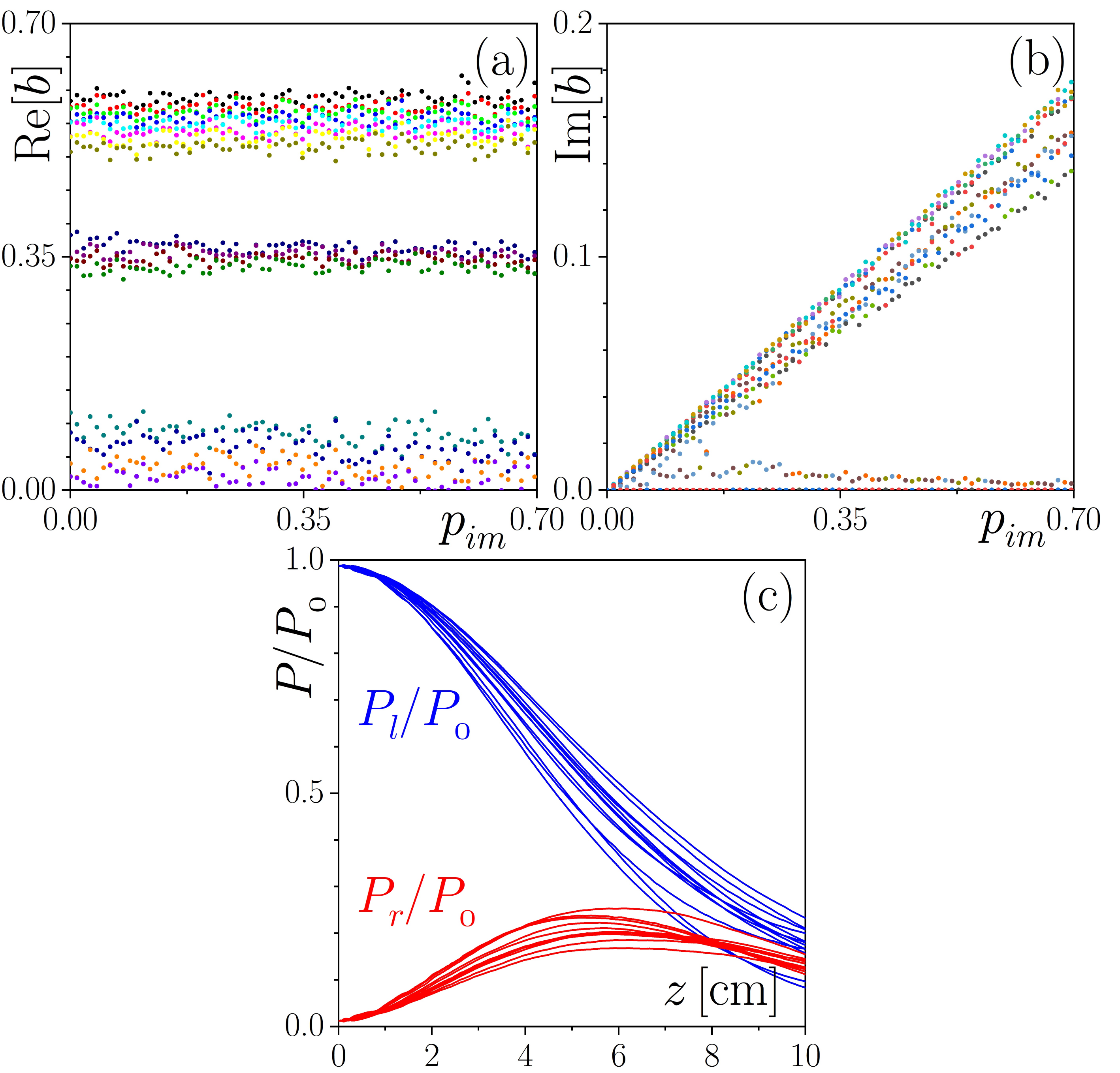}
\caption{
(a) Real and (b) imaginary parts of the propagation constant $b$ of the modes supported by the TT array as functions of $p_{\rm im}$ for disorder strength $\delta p_{\rm re}=0.02$ (maximum refractive index deviation is $1.1\times10^{-5}$) and $\delta_x=0.1$ ($1$~$\mu \rm m$). (c) Powers concentrated in the left, $P_l$, and right, $P_r$, arrays as functions of the propagation distance $z$ at $p_{\rm im}=0.1$ for the different realizations of disorder with $\delta p_{\rm re}=0.02$, $\delta p_{\rm im}=0.1$ (maximum deviation of loss coefficient is $0.04$~$\rm cm^{-1}$), and $\delta_x=0.1$. Here $h=39$~$\mu \rm m$.}
\label{fig7}
\end{figure}

To examine the robustness of the Zeno effect in our topological system, we introduce the diagonal and off-diagonal disorder in the array. We numerically calculated the spectrum of the TT structure, where the depth of each waveguide randomly varies in the interval $[p_{\rm re}(1-\delta p_{\rm re})$, $p_{\rm re}(1+\delta p_{\rm re})]$, and the spatial position of the waveguides is shifted by a random value from the range $[-\delta_x,\delta_x]$. {We consider values of $\delta p_\textrm{re}=0.02$ that correspond to typical level of fluctuations of the refractive index in the waveguides that may arise due to the fluctuations of power of writing laser and $\delta_x=0.1$ that dramatically exceeds the error of our positioning system $\sim 0.01~\mu \textrm{m}$.} As can be seen from the eigenvalue spectrum in Fig.~\ref{fig7}(a) and~(b), the presence of the disorder broadens the bulk bands and slightly shifts the propagation constant of the topological levels for all values of $p_{\rm im}$. Nevertheless, these levels remain in the topological gap and characteristic behavior with the different evolution of $\textrm{Im}(b)$ with the increase of $p_{\rm im}$ for two topological modes (a signature that Zeno effect is possible) in the gap persists. We also confirmed the robustness of the effect by direct propagation of excitation in a disordered TT structure, where we also added a disorder in the losses in the right array so that not only the real part of the refractive index, but also $p_{\rm im}$ varies randomly and uniformly over the interval $[p_{\rm im}(1-\delta p_{\rm im}),p_{\rm im}(1+\delta p_{\rm im})]$. The dependencies of powers concentrated in the left and right arrays on propagation distance $z$ for the different disorder realizations are shown in Fig.~\ref{fig7}(c). As one can see, there are only small deviations in these dependencies from realization to realization, but the overall behavior remains the same as for the unperturbed array.

\section{Conclusions}

In conclusion, we experimentally observed the occurrence of the macroscopic optical Zeno effect in a topological insulator, as the counter-intuitive increase of the output energy with an increase of the absorption of one of the arrays constituting the structure. Importantly, we have shown that the presence of the Zeno effect depends on the topological phase of the subsystems. Our observations can be readily extended to the photonic valley Hall schemes, Floquet topological insulators and two-dimensional schemes based on the waveguide arrays. From a practical viewpoint, our observation opens the possibility to the realization of topological switching schemes based on the method of controlling artificial losses in a waveguide system~\cite{GuoSalamo-09,RuterMakris-10,RegensburgerBersch-12,KremerBiesenthal-19}. {{Importantly, the effect reported here may also occur in other systems beyond waveguide arrays, highlighting the potential for controlling topological states in other systems such as metal-dielectric multilayers employed to generate topologically protected plasmonic surface modes (see, e.g.,~\cite{DengChen-16(1),DengChen-16(2)}}).} Also, our work provides tools to experimentally address the impact of nonlinear Zeno effect on switching (see, e.g.,~\cite{AbdullaevKonotopShchesnovich-11}) and the possibility to control topological solitons by dissipation.


\section*{Methods}

\subsection{Numerical simulations}

\begin{figure}[t]
\centering
\includegraphics[width=1.0\linewidth]{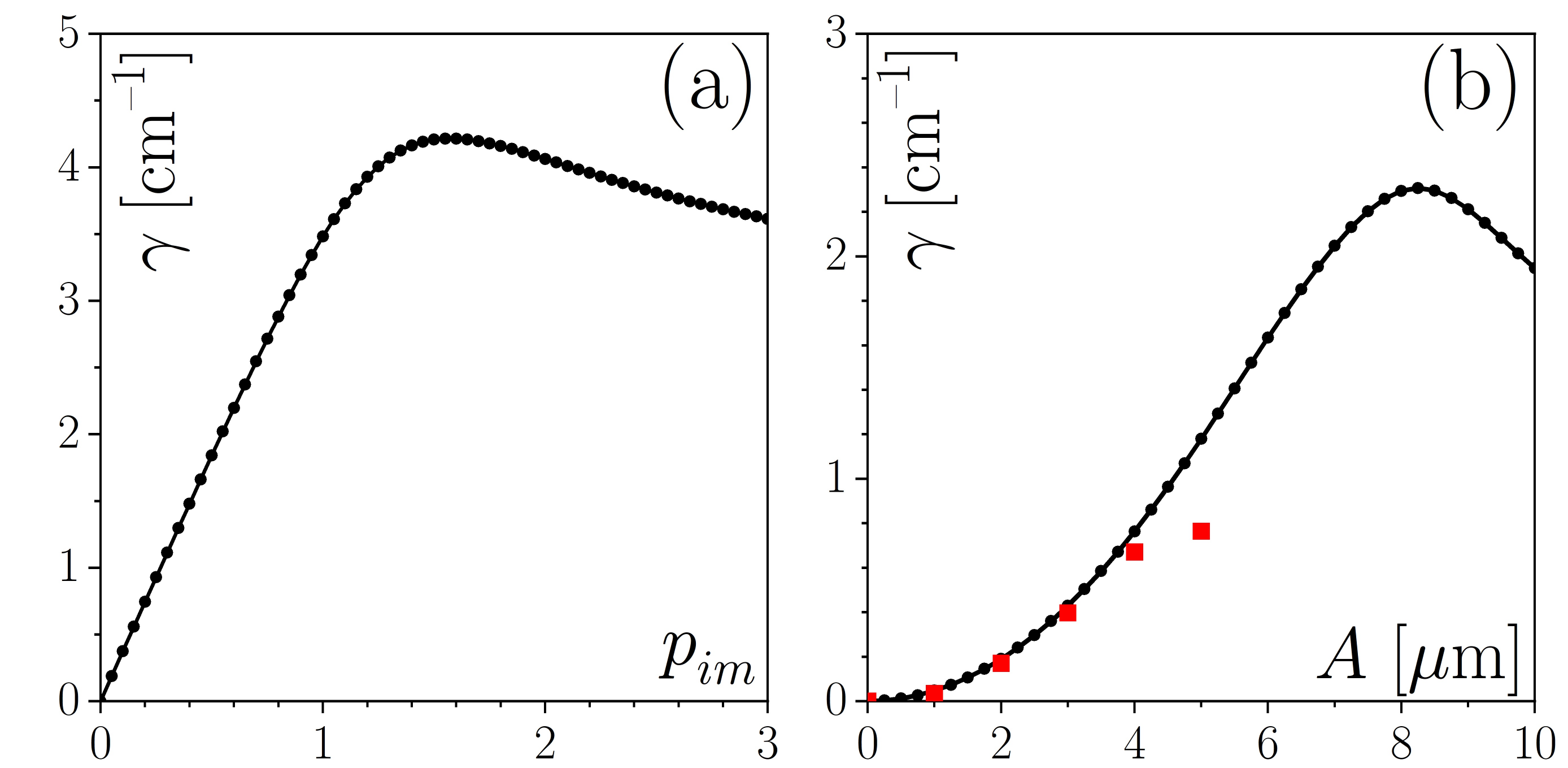}
\caption{
Dependence of the loss factor $\gamma$~($\rm cm^{-1}$) on the imaginary depth $p_{\rm im}$ of the potential (a) and on the amplitude $A_l$~($\mu \rm m$) of the waveguide wiggling (b). Red dots correspond to the experimentally measured loss factors.}
\label{fig8}
\end{figure}

For the calculation of the spectrum and modes of the system, we use the standard finite difference method for the time-independent two-dimensional Schr\"{o}dinger equation. The slip-step fast Fourier method is used for modeling light propagation. For theoretical analysis, we vary the parameter $p_{\rm im}$ (depth of the imaginary part of the potential), resulting in the loss factor $\gamma=(1/z)\, \log{[P_0/P]}$, where $z$ is the propagation distance, $P_0$ and $P$ are the input and output powers $P=\iint |\psi|^2 dxdy$, respectively. The dependence of actual loss factor $\gamma$ (in $\rm cm^{-1}$) on the parameter $p_{\rm im}$ for a single waveguide is shown in Fig.~\ref{fig8}(a) for the waveguide parameters and wavelength used in our experiments. The initial condition that was used to calculate this dependence is a single-channel mode whose width is close to the experimentally measured mode width. In this work, we consider $p_{\rm im}$ values in the range of $0$ to $1.55$, that is, only the monotonic part of the curve.

\subsection{Experiments}

For the experimental observation of the macroscopic optical Zeno effect, we use the SSH arrays with $20$ waveguides ($5$ dimers in each array) inscribed inside $10$~$\rm cm$-long fused silica glass samples by focused (using an aspheric lens with $\rm NA = 0.3$) femtosecond laser pulses (wavelength $515$~$\rm nm$, pulse duration $280$~$\rm fs$, pulse energy $340$~$\rm nJ$, repetition rate $1$~$\rm MHz$) under the surface of sample at the depth of $800$~$\mu\rm m$. During the inscription process, the sample was translated relative to the focus at the constant velocity of $1$ $\rm mm/s$ using a high-precision air-bearing positioner system (Aerotech), resulting in the inscription of sets of parallel waveguides with the controllable spacing between them. The average refractive index modification in such waveguides is about $\delta n\sim 5.3\times10^{-4}$, i.e., they are single-mode at $800$~$\rm nm$ with the mode field diameter $d_x \times d_y$ $\sim 15.4\times24.0$~$\mu\rm m$.

In our experiments, extra losses were introduced using rapid waveguide wiggling in the direction orthogonal to the array. The light is most strongly radiated away from the waveguides where the curvature is the largest. The dependence of the loss factor $\gamma$ on the oscillation amplitude for a single waveguide is shown in Fig.~\ref{fig8}(b), where black dots correspond to the numerical calculations and red dots to the experiment. {{When conducting numerical estimates, it is crucial to consider the ellipticity of the waveguides, as the oscillations along different axes of the ellipse can result in different levels of losses. In our case oscillations of waveguides were always along the longer axes of the ellipses that can be easily controlled in fs-laser writing technique.}} For amplitudes larger than $5$~$\mu\rm m$, the losses become significant and the output power is comparable to the background illumination. Therefore, for the high amplitudes, we use theoretical calibration. It should be noted that the connection between the theoretical parameter $p_{\rm im}$ and $A_l$ is nonlinear, therefore we use the calibration results from Fig.~\ref{fig8} and operate with the losses $\gamma$.


~




\textbf{Acknowledgments.}
S.A.Z. acknowledges support of the Interdisciplinary Scientific and Educational School of Moscow State University ``Photonic and Quantum Technologies. Digital Medicine''.


\textbf{Funding.}
Funding of this study by Russian Science Foundation under grant 21-12-00096, and partially by research project FFUU-2021-0003 of the Institute of Spectroscopy of the Russian Academy of Sciences is acknowledged.
V.V.K. was supported by the Portuguese Foundation for Science and Technology (FCT) under Contract UIDB/00618/2020.


~

\appendix

\section{Two-mode model}

\begin{figure}[h]
\centering
\includegraphics[width=1\linewidth]{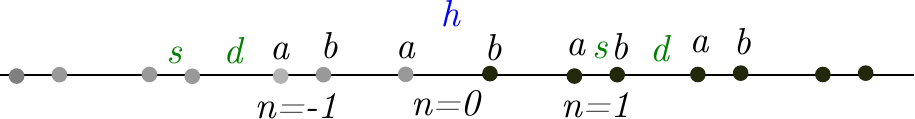}
\caption{Site numbering in the TT configuration of two SSH arrays.}
\label{fig1S}
\end{figure}

Here we derive the two-mode model (5) 
for the TT configuration for which it is more convenient to use numbering of the cells (i.e. below $n$ refers to a number of a cell rather than a waveguide, as in the main text) and treat the Su–Schrieffer–Heeger (SSH) arrays as the coupled dimers referring to the left and right waveguide in each dimer $a$ and $b$ waveguide (Fig.~\ref{fig1S}).

Then, the optical potential can be written in the form
 \begin{align}
		\label{Potential-short}
		V(\br) =
		\sum_{n=-\infty}^{\infty}\left[  V_0(\br-\br_n^{a} )+V_0 (\br- \br_n^{b}  )\right],
	\end{align}

\noindent where to shorten the notations we will use $V_0(x,y)\equiv V_0(\br)$ and define 
\begin{align*}
\br_n^{a}&={\bf i}\begin{cases}
	 n \ell   & n\leq 0
	 \\
	 n\ell+h-s  & n\geq 1
\end{cases}	, 
\quad
\br_n^{b}&={\bf i}\begin{cases}
	 n \ell +s & n\leq 0
	 \\
	 n\ell+h & n\geq 1
\end{cases} ,
\end{align*}

\noindent $\ell=s+d$, and  ${\bf i}$ is the unit vector along $x-$axis. By analogy, the absorption takes the form:
\begin{align}
\label{loss}
	W(\br)
		=& \sum_{n=1}^{\infty}W_0(\br-\br_n^{a})+ \sum_{n=0}^{\infty} W_0(\br-\br_n^{b} ) .
\end{align}

One can then search for the solution of Eq.~(1) 
in the tight-binding approximation in the form 
\begin{align}
 	 \psi(\br,z)=e^{i\beta z}\sum_{n=-\infty}^\infty \left[ a_n(z)\phi(\br-\br_n^a)+b_n(z)\phi(\br-\br_n^b)\right].
\end{align}

To compute $H\psi$ we use this ansatz and neglect all integrals except those involving the next-neighbor interactions, i.e. integrals of the types   
 \begin{align*}
 	\label{neglect1}
	 &\int_{\mathbb{R}^2}|\phi(\br)|^2V_0(\br \pm \ell {\bf i})d^2r,\\
	 &\int_{\mathbb{R}^2}|\phi(\br)|^2W_0(\br \pm \ell {\bf i})d^2r,\\
	 &\int_{\mathbb{R}^2}\phi^*(\br)V_0(\br \pm \ell{\bf i})\phi(\br\pm \ell{\bf i})d^2r,\\
	 &\int_{\mathbb{R}^2}\phi^*(\br)W_0(\br \pm \ell {\bf i})\phi(\br\pm \ell{\bf i})d^2r,
\end{align*}

\noindent as well as the interactions beyond the nearest neighbors. Then, we will obtain the following system of equations:
\begin{align}
	\begin{cases}
		\displaystyle -i\frac{da_n}{dz} =\varepsilon  a_n +v_db_{n-1}+v_sb_n& n\leq -1\\[2mm]
		\displaystyle -i\frac{db_n}{dz} =\varepsilon  b_n +v_sa_{n}+v_da_{n+1}& n\leq -1\\[2mm]
		\displaystyle -i\frac{da_0}{dz} =(\varepsilon_0+iw_a)  a_0 +v_db_{-1}+(v_h+iw_h )b_0& n=0\\[2mm]
		\displaystyle -i\frac{db_0}{dz} =(\varepsilon_0+iw_b) b_0 +(v_h+iw_h) a_0 +(v_d +iw_d)a_1 & n=0
		\\[2mm]
		\displaystyle -i\frac{da_n}{dz} =(\varepsilon+iw)  a_n +(v_d+iw_d)b_{n-1}+(v_s+iw_s)b_n& n\geq 1\\[2mm]
		\displaystyle -i\frac{db_n}{dz} =(\varepsilon+iw)  b_n +(v_s+iw_s)a_{n}+(v_d+iw_d)a_{n+1}& n\geq 1
	\end{cases},
\end{align} 

\noindent where 
\begin{align*}
	\varepsilon=&
	\int_{\mathbb{R}^2}|\phi(\br)|^2[V_0(\br+d{\bf i})+V_0(\br-s{\bf i})]d^2r
	\\
\varepsilon_0=&\int_{\mathbb{R}^2}|\phi(\br)|^2[V_0(\br+d{\bf i})+V_0(\br-h{\bf i})]d^2r,
\\
	w=& 
	\int_{\mathbb{R}^2}|\phi(\br)|^2[W_0(\br+d{\bf i})+W_0(\br)+V_0(\br-s{\bf i})]d^2r,
		\\
	w_b=& \int_{\mathbb{R}^2}|\phi(\br)|^2[W_0(\br)+W_0(\br-d{\bf i})]d^2r,
	\\
	w_a=& \int_{\mathbb{R}^2}|\phi(\br)|^2W_0(\br-h{\bf i})d^2r,
	\\
	w_s =&2\int_{\mathbb{R}^2}\phi^*(\br) W_0(\br - s{\bf i})\phi(\br-s{\bf i})d^2r,
	\\
	w_d =&2\int_{\mathbb{R}^2}\phi^*(\br) W_0(\br - d{\bf i})\phi(\br-d{\bf i})d^2r,
 	\\
	w_h =&\int_{\mathbb{R}^2}\phi^*(\br)W_0(\br - h{\bf i})\phi(\br-h{\bf i})d^2r,
\\
	 v_s =&\int_{\mathbb{R}^2}\phi^*(\br)V_0(\br - s{\bf i})\phi(\br-s{\bf i})d^2r,
	 \\
	 v_d =&\int_{\mathbb{R}^2}\phi^*(\br)V_0(\br - d{\bf i})\phi(\br-d{\bf i})d^2r,
     \\
 	 v_h =&\int_{\mathbb{R}^2}\phi^*(\br) V_0(\br - h{\bf i})\phi(\br-h{\bf i})d^2r.
\end{align*}
 
Finally, to obtain the two-mode model (5) in the main text  we consider the cell $n=0$ and  allow $d\to\infty$ holding $h$ finite (i.e., we assume that the fields are localized only in two central waveguides at the interface), and make the ansatz  
\begin{align}
	a_0=A_0e^{i\varepsilon_0 z},\qquad b_0=B_0e^{i\varepsilon_0 z}.
\end{align}

{{
\section{Spectrum and dynamics of beams at the interface of two arrays in the trivial phase}

Next, we consider a system that consists of two SHH arrays in the trivial phase
\begin{align}
\label{}
    V(\br) =
    \sum_{n=-\infty}^{\infty}\left[  V_0(\br-\br_n^{a} )+V_0 (\br-\br_n^{b}  )\right],
\end{align}
where 
\begin{align*}
\br_n^{a}&={\bf i}\begin{cases}
	 n \ell   & n\leq 0
	 \\
	 n\ell+h-d  & n\geq 1
\end{cases}	, 
\quad
\br_n^{b}&={\bf i}\begin{cases}
	 n \ell +d & n\leq 0
	 \\
	 n\ell+h & n\geq 1
\end{cases} ,
\end{align*}

\begin{figure}[h]
\centering
\includegraphics[width=1\linewidth]{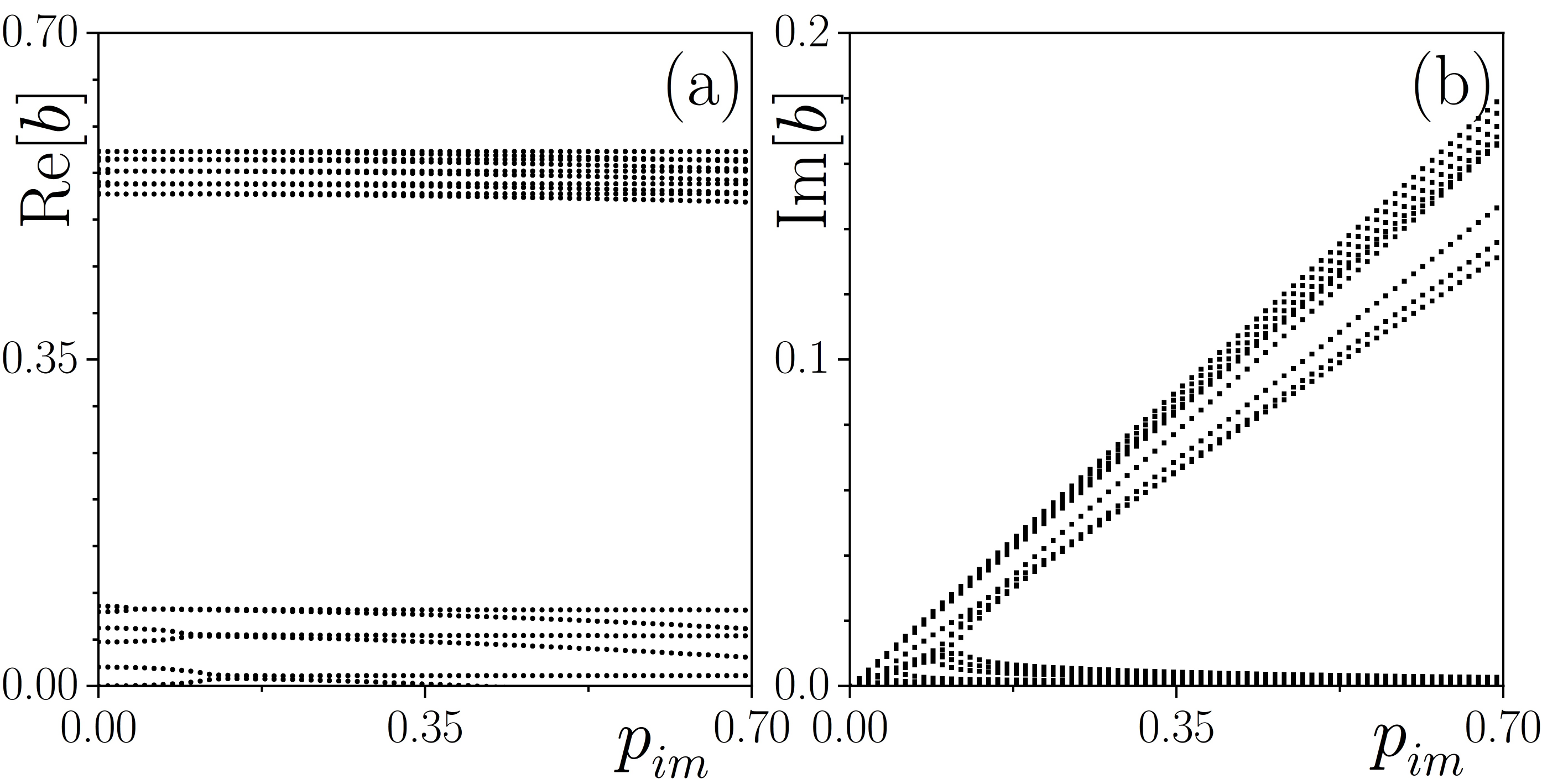}
\caption{Real (a) and imaginary (b) parts of the propagation constant $b$ of the modes supported by the array with two trivial SHH arrays as functions of $p_{\rm im}$.}
\label{fig2S}
\end{figure}

\begin{figure}[h]
\centering
\includegraphics[width=0.9\linewidth]{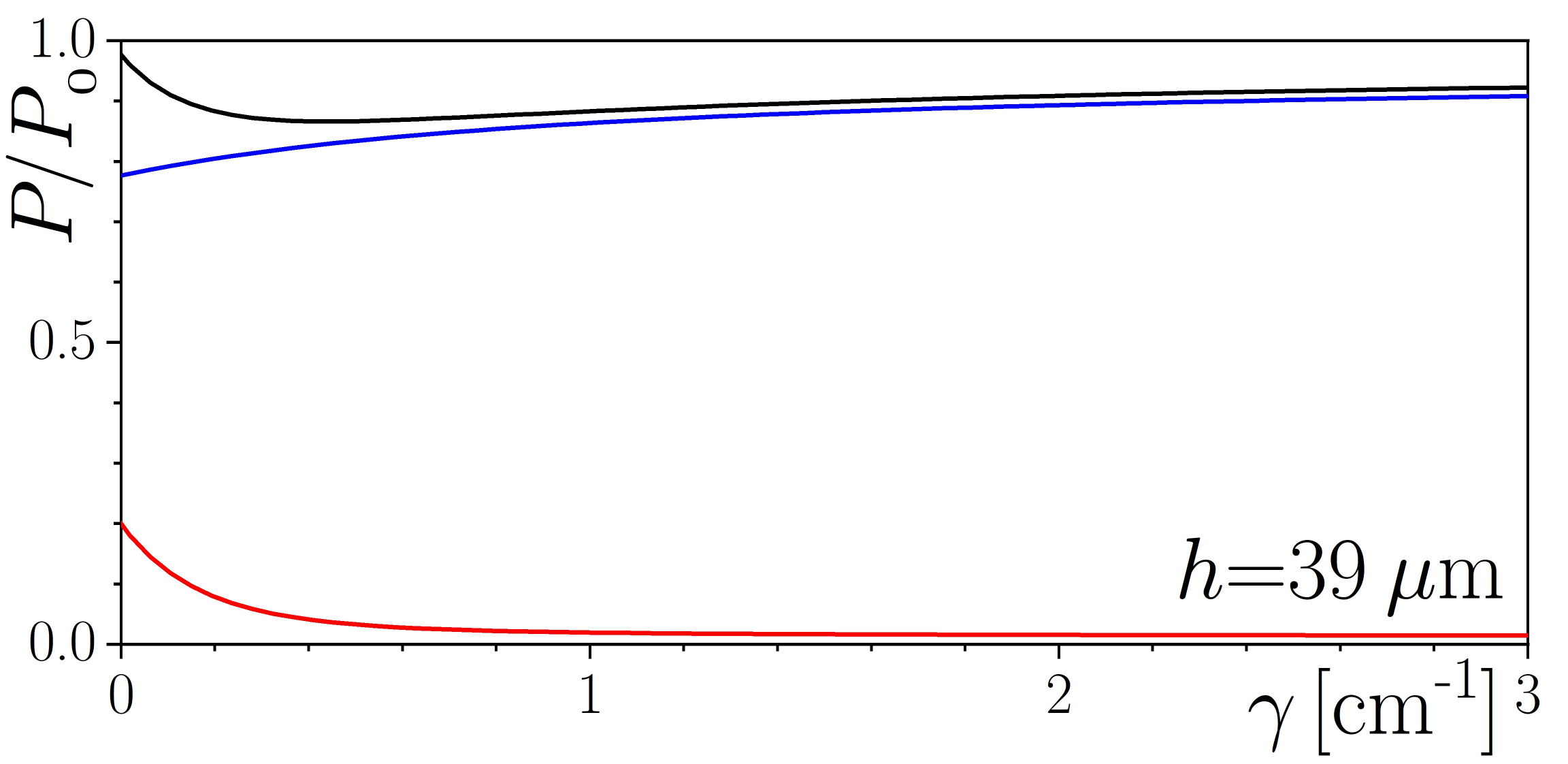}
\caption{Dependencies of the output power $P$ and powers in the left $P_l$ and right $P_r$ arrays on the loss factor $\gamma$ for the distance $h=39~\mu \textrm{m}$ between two arrays in the trivial phase. The power is normalized to the output power of the sample with straight waveguides $P_0$ (without additional losses in the right array). Black, blue, and red curves correspond to the total output power, and the output power in the left (lossless) and right (lossy) arrays, respectively.}
\label{fig3S}
\end{figure}

We assume also that the left array is transparent, while the right arm is lossy [see Eq.~(\ref{loss})]. Parameter $p_{\rm im}$ controls the amount of losses. We numerically calculated the transformation of the spectrum for this configuration with an increase of $p_{\rm im}$. The dependence of the real and imaginary parts of $b$ on $p_{\rm im}$ is shown in Figs.~\ref{fig2S}(a) and~(b) for the parameters $p_{\rm re}=4.7$, $d_l=d_r= 1.5$, $s_l=s_r = 3.3$. As can be seen, there are no branches that belong to the gap that indicates that the structure is in the trivial phase and all modes of this system are delocalized.
Thus, all modes of a topologically trivial (NN) array do not belong to a Hilbert space. Hence one cannot select a subspace that can emulate the effect of frequent measurements while being affected by dissipation. Consequently, such an array is unable to replicate the decay of a quantum system, unlike the case of TT arrays where topological states form a Hilbert space. Although we observe similarities in the behavior of $\textrm{Im}(b)$ in both cases, the topologically trivial array cannot be considered for demonstration of the Zeno phenomenon. When performing numerical calculations with finite arrays, this limitation becomes apparent. In the trivial phase, modes that exhibit non-monotonic behavior of $\textrm{Im}(b)$ versus $p_{im}$ are extended and possess an energy density that is nearly negligible. Consequently, these modes will have an insignificant amplitude at the output of the array, so their Zeno dynamics cannot be observed in such cases.

To illustrate this, we calculated the dependence of the output total power $P$ as well as the output power concentrated in the left $P_l$  and right $P_r$ arrays on the strength of losses in the right array [see Fig.~\ref{fig3S}]. As in the main text, we excited only one waveguide of the left array at the interface. Since this NN structure has no localized modes, light strongly spreads across the arrays, especially in the left one. Only a small fraction of power couples into the right array. As a result, output power only slightly changes with $\gamma$ in contrast to strong variations in transmission observed in the TT structures in the main text.
}}




\end{document}